# Entanglement distribution over 300 km of fiber

Takahiro Inagaki,[1,*] Nobuyuki Matsuda,[1] Osamu Tadanaga,[2]
Masaki Asobe,[2] and Hiroki Takesue[1]

[1]*NTT Basic Research Laboratories, NTT Corporation, Atsugi, Kanagawa, 243-0198, Japan*
[2]*NTT Photonics Laboratories, NTT Corporation, Atsugi, Kanagawa, 243-0198, Japan*
[*] *inagaki.takahiro@lab.ntt.co.jp*

**Abstract:** We report the distribution of time-bin entangled photon pairs over 300 km of optical fiber. We realized this by using a high-speed and high signal-to-noise ratio entanglement generation/evaluation setup that consists of periodically poled lithium niobate waveguides and superconducting single photon detectors. The observed two-photon interference fringes exhibited a visibility of 84 %. We confirmed the violation of Bell's inequality by 2.9 standard deviations.

## 1. Introduction

The distribution of quantum information over long distances is important if we are to realize large-scale quantum communication networks [1]. Several record-setting quantum communication experiments have already been reported, including quantum key distribution (QKD) over 260 km of fiber [2], entanglement distribution over 144 km in free space [3] and over 200 km of fiber [4], and quantum teleportation over 97 km [5] and 144 km [6] in free space. Long-distance entanglement distribution is particularly important for quantum communications and quantum repeater systems [7-11]. Numerous experiments have been reported on entanglement distribution in the 1.5-μm wavelength band (telecom band) over optical fiber [4, 12-18]. Although a standard single mode fiber has a reasonable loss of a 0.2 dB/km, the coincidence count between distributed entangled photon pairs decreases exponentially as the transmission distance increases. The distribution distance limitation depends on the signal-to-noise ratio of the systems, and so a low-noise experimental setup is required for a long-distance entanglement distribution.

In this paper, we demonstrate the distribution of time-bin entangled photon pairs over 300 km of optical fiber. We realize this demonstration by using a high-speed and high signal-to-noise ratio entanglement generation/evaluation setup. We can generate entangled photon pairs with low background noise in the 1.5-μm wavelength band by using the spontaneous parametric down conversion (SPDC) process in periodically poled lithium niobate (PPLN) [19, 20] waveguides with a high-speed pulsed pump source. The superconducting single photon detectors (SSPDs) [21] contribute to the highly efficient detection of entangled photon pairs with low-noise. After separating the signal and idler photons by 300 km of fiber, we observed high-visibility two-photon interference fringes without subtracting accidental coincidences. Furthermore, we confirmed the violation of Bell's inequality in accordance with the Clauser, Horne, Shimony, and Holt (CHSH) inequality [22].

## 2. Time-bin entanglement generation and measurement.

Quantum information encoded in different temporal modes of photon, which is called a time-bin qubit, is generally used in practical quantum communications over optical fiber [23]. In contrast to the polarization state, the relative phase between the temporal modes in a time-bin qubit is stable against refractive index and birefringence fluctuations in the optical fiber. Time-bin entanglement is generated through the SPDC process in a non-linear crystal pumped by temporally separated coherent pulses. To increase the operation frequency so that we can use the time domain more efficiently, we generate high-dimensional time-bin entangled photon pairs [24] by employing the SPDC process in a PPLN waveguide pumped by sequential pulses. For a very low mean photon pair number, the whole state wave function of a photon pair can be written approximately as:

$$|\psi\rangle = \frac{1}{\sqrt{N}} \sum_{k=1}^{N} |k\rangle_s |k\rangle_i , \qquad (1)$$

where $|k\rangle_z$ denotes a state in which a single photon is found at the $k$th time slot in the mode $z$ (=$s$: signal, $i$: idler). $N$ is the number of pulses in which the phase coherence of the pump pulses is preserved. After spectral separation, the signal and idler photons are launched into two different long-distance fibers. We evaluate the degree of entanglement between the signal and idler photon pairs by observing two-photon interference in a Franson-type experiment [25, 26]. The distributed photons in each channel are measured with an $n$-bit ($1 \leq n \leq N$) delayed Mach–Zehnder interferometer (MZI) followed by an SSPD. The $n$-bit delayed MZI transforms the $|k\rangle_z$ state into the $|k\rangle_z + \exp(i\theta_z)|k + n\rangle_z$ state (unnormalized), where $\theta_z$ is the relative phase between the two divided paths in the MZI. Then, Eq. (1) is converted to the following equation.

$$|\psi\rangle \Rightarrow \sum_{k=1}^{n} |k\rangle_s |k\rangle_i + \sum_{k=n+1}^{N} \left(1 + e^{i(\theta_s + \theta_i)}\right) |k\rangle_s |k\rangle_i + \sum_{k=N+1}^{N+n} e^{i(\theta_s + \theta_i)} |k\rangle_s |k\rangle_i , \qquad (2)$$

where only the terms that contribute to the coincidence counts are extracted without normalization for simplicity. Under the condition of $N \gg n$, the coincidence count rate between the signal and idler photons is approximately proportional to $1 + V\cos(\theta_s + \theta_i)$, where $V$ is the visibility of the two-photon interference. Consequently, the coincidence count rate changes depending on the phases of the MZIs placed in distant locations.

To confirm the violation of Bell's inequality, we measure the $S$ value for the CHSH inequality [22, 27, 28], $|S| \leq 2$ for any local realistic theory, where

$$S = E(d_s, d_i) + E(d_s, d'_i) + E(d'_s, d_i) - E(d'_s, d'_i). \qquad (3)$$

$d_z$, $d'_z$ ($z = s, i$) denote arbitrary values of phases $\theta_z$ in interferometers, and $E(\theta_s, \theta_i)$ is defined as:

$$E(\theta_s, \theta_i) := \frac{R(\theta_s, \theta_i) - R(\theta_s, \theta_i + \pi) - R(\theta_s + \pi, \theta_i) + R(\theta_s + \pi, \theta_i + \pi)}{R(\theta_s, \theta_i) + R(\theta_s, \theta_i + \pi) + R(\theta_s + \pi, \theta_i) + R(\theta_s + \pi, \theta_i + \pi)}. \qquad (4)$$

Here $R(\theta_s, \theta_i)$ is the coincidence count rate of a two-photon interference measurement when we set the phases of the interferometer for the signal and idler photon at $\theta_s$ and $\theta_i$, respectively. The measured value of $|S| > 2$ indicates the violation of Bell's inequality. A maximum value of $|S| = 2\sqrt{2}$ is predicted by quantum mechanics.

## 3. Experimental setup

Figure 1 shows the experimental setup. A continuous wave light (1551 nm) from an external-cavity diode laser was modulated into a sequence of 72-ps pulses at a temporal interval of 500 ps. The pulses were amplified by an erbium-doped fiber amplifier (EDFA) and filtered by a fiber Bragg grating (FBG) to suppress the amplified spontaneous emission noise from the EDFA. The pulses were polarization-controlled and then launched into the first PPLN waveguide, where a sequence of pulses was generated at a wavelength of 775.5 nm by the second harmonic generation process. The output light from the first PPLN waveguide was input into filters that transmitted the 775.5 nm light while eliminating the remaining 1551 nm light. The generated pulses (775.5 nm) acted as the pump light for the SPDC in the second PPLN waveguide, and high-dimensional time-bin entangled photon pairs were generated whose states are shown by Eq. (1). $N$ of the high-dimensional time-bin entangled state was estimated from the coherence time and the temporal interval of the pump pulses. The coherence time of the laser was approximately 10 μs, and the temporal interval of the pump pulses was 500 ps. Thus we estimated $N \approx 20,000$. According to the energy conservation law, the frequencies of the pump $f_p$, signal $f_s$, and idler $f_i$ had a relationship of $f_p = f_s + f_i$. The loss in the PPLN waveguide for the SPDC process was estimated to be about 3 dB, which includes the propagation loss of the waveguide and the coupling loss at the output with single mode fiber. The output from the second PPLN waveguide was input into filters that transmitted the 1.5 μm light while eliminating the remaining 775.5 nm light. The excess loss of this filter was 0.19 dB for 1551 nm photons. The generated photon pairs were separated into signal (1547 nm) and idler (1555 nm) photons by a wavelength selective filter. The spectral width of this filter was 100 GHz, and its excess loss was 0.69 dB at 1547 nm and 0.61 dB at 1555 nm. Then, each photon was transmitted over a 150 km dispersion shifted fiber (DSF). The total distance traveled by the separated signal and idler photons was 300 km, and the total loss for the entire transmission exceeded 64 dB. Note that the total loss included only the loss of the quantum channel (*i.e.* the DSFs and fiber connections), and did not include the losses of other optics components, such as filters and interferometers. The distributed signal and idler photons were launched into MZIs fabricated based on planar lightwave circuit (PLC) technology. The propagation time difference between the two paths of the MZI was 1 ns, and thus it worked as a 2-bit delay for 2 GHz repetition pulses. The temperature and phase of the MZI have a simple linear relationship. The phase of the MZI, $\theta_z$, was tuned by $2\pi$ by changing the MZI temperature by 0.74 °C. An output port of each MZI was connected to an SSPD (SCONTEL) through a polarization controller (PC). The detection efficiencies for the signal and idler photons were 15 and 20 %, and dark count rates were 10 and 15 Hz, respectively. As a single photon detection system that consists of the SSPD and discriminators, the jitter and dead time were 50 ps and 50 ns, respectively. Output signals from the SSPDs were input into a time interval analyzer (TIA) as the start and stop signals for coincidence detection. The time resolution of the TIA was 9.8 ps, and the coincidence counts were collected in a 300 ps time window.

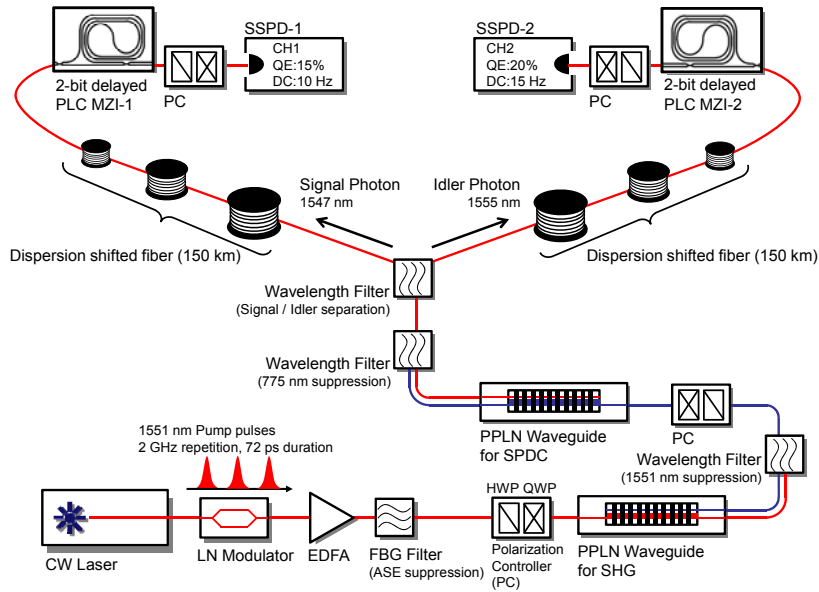

Fig.1. Experimental setup for distribution of time-bin entangled photon pairs over 300 km of optical fiber. EDFA: Erbium doped fiber amplifier, FBG filter: fiber Bragg grating to suppress amplified spontaneous emission from the EDFA, PC: polarization controller, PPLN: periodically poled lithium niobate waveguide, PLC MZI: Mach–Zehnder interferometer fabricated based on a planar lightwave circuit.

## 4. Signal-to-noise evaluation

In our experimental setup, low-noise entangled photon pairs were generated by the SPDC process in the PPLN waveguide, which made it possible to isolate the generated photon pairs (1547 and 1555 nm) from the pump light (775.5 nm) efficiently with wavelength filters. The low dark count probability of the SSPD of about $1 \times 10^{-9}$ within the 100 ps time window contributed to the low-noise detection of the entangled photon pairs. To evaluate the signal-to-noise ratio $R_{SN}$ of our entanglement generation/evaluation system, we removed the DSFs and PLC-MZIs from the setup and measured the coincidence-to-accidental ratio, which is defined by $R_c / R_{acc}$, where $R_c$ and $R_{acc}$ are coincidence and accidental count rates, respectively. The signal-to-noise ratio can be calculated by the following equation.

$$R_{SN} = \frac{R_c}{R_{acc}} = \frac{\mu \alpha_s \alpha_i + (\mu \alpha_s + d_s)(\mu \alpha_i + d_i)}{(\mu \alpha_s + d_s)(\mu \alpha_i + d_i)}, \quad (5)$$

where $\mu$, $\alpha_z$, and $d_z$ are the average number of generated photon pairs per pulse, the transmittance and the dark count rate for the mode $z$ (=$s$: signal, $i$: idler), respectively. Figure 2 shows the obtained $R_{SN}$ as a function of $\mu$. The dashed blue line shows the theoretical $R_{SN}$ calculated on the assumption of no dark counts. The solid red line shows the $R_{SN}$ calculated by Eq. (5) with the experimental parameters. The black squares are the experimental results. We used 1 GHz repetition pump pulses, and collected the coincidence counts within a time-window of 600 ps. The large error bar in the $R_{SN}$ measurement result was because there were too few accidental coincidences, thanks to the extremely small dark count rate of the SSPDs. For example, at a mean photon number $\mu = 3 \times 10^{-7}$, we observed 4608 coincidences at the matched time slot in 1.5 hours, while we observed only 21 accidental coincidences in the 4000 unmatched time slots in the same measurement time. We obtained the maximum $R_{SN}$ over 800,000 at $\mu = 3 \times 10^{-7}$, which is the highest coincidence-to-accidental ratio yet reported. Thus, we confirmed that our setup has a very high signal-to-noise ratio.

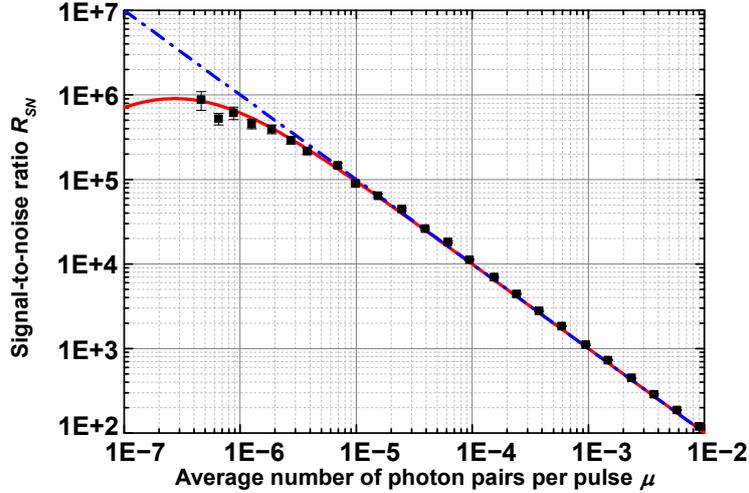

Fig.2. Signal-to-noise ratio of experimental setup. The signal-to-noise ratio $R_{SN}$ of our entanglement generation/evaluation system was obtained from the coincidence and accidental count rates. The error bars were calculated by using the square root of the coincidence and accident counts.

## 5. Two-photon interference experiment

We then distributed time-bin entangled photon pairs over 300 km of DSF and observed the two-photon interference. First, we fixed $\theta_s$ by setting the temperature of MZI-1 at 15.35 °C and swept the temperature of MZI-2 in 0.1 °C steps while measuring the coincidence counts. For this measurement, we set the pump power for the SPDC process at 250 μW to give $\mu = 0.1$.

We collected one hour's worth of data for each temperature setting. The corrected coincidence counts, which are raw data without the subtraction of accidental counts, are shown by the circles in Fig. 3. The visibility of the fitted curve was 86.1 ± 6.8 %. We then changed $\theta_s$ with the temperature of MZI-1 at 15.54 °C and observed another fringe (squares), whose visibility was 83.7 ± 9.1 %. The deviation of some data points, which were larger than the errors that take account of the statistical fluctuation, were caused by the fluctuations of the setup and the laboratory environment such as variations in room temperature, and the experimental fluctuation of various parameters including the pump power, and the detection efficiencies. Figure 4 shows the TIA histograms around the coincidence peaks, which correspond to the coincidence count data (circles) in Fig. 3 for each MZI temperature setting. We collected the coincidence counts were collected within a 300 ps time window over a period of one hour. The temporal positions of the coincidence peaks were shifted slowly because of the fluctuations in fiber length caused by the unstable room temperature. Thus, we had to trace the coincidence peaks by shifting the time window. If we use a lower $\mu$, we need a longer measurement time to collect sufficient coincidence counts, and the fluctuation of the temporal positions of the coincidence peaks can be larger than 500 ps, which is the time interval of the photon pairs in our experiment. Thus we limit the measurement time to an hour and set $\mu$ at 0.1 so that we can achieve a coincidence count of about 100 counts / hour under maximized MZI phase conditions.

We then performed an $S$ value measurement for the CHSH inequality. First, we searched for a phase setting where the coincidence rate was maximized, and we defined these phases as $\theta_{s0}$ and $\theta_{i0}$. We used these phases to decide the measurement parameters in eq. (3) as $d_s = \theta_{s0}$, $d'_s = \theta_{s0} + \pi / 2$, $d_i = \theta_{i0} + \pi / 4$, and $d'_i = \theta_{i0} - \pi / 4$. We obtained 16 values of $R(\theta_s, \theta_i)$ for the calculation of the $S$ value over 16 hours. To improve the accuracy of the obtained $S$ value, we repeated the same experiments three times with the same measurement parameters. Table 1 shows the total coincident count for each phase setup and the calculated values of $E(\theta_s, \theta_i)$. There were two possible sources of the phase error: the MZI phase uncertainty caused by the finite accuracy of the temperature controller, and the phase drift due to the fluctuation of the pump laser frequency. The accuracy of our temperature control was ± 0.01 °C, which resulted in an MZI phase error of ± 84 mrad. We also measured the stability of the pump laser frequency and obtained a frequency drift of < 570 kHz/hour. This means that the error due to the pump frequency drift was < 3.6 mrad/hour, which is much smaller than the MZI phase error. In addition, in the $S$ value measurement, we calibrated the phases of $\theta_{s0}$ and $\theta_{i0}$ every 8 hours so that we could avoid the accumulation of phase setting uncertainties from the frequency drift. Thus we consider that the MZI phase error was dominant in the phase measurement error. The estimated errors from the coincidence count statistics and the uncertainties of MZI phase settings for each set of measurement data are shown in Table 1. As a result, we found that $S$ = 2.41 ± 0.14 leading to the violation of 2.9 standard deviations.

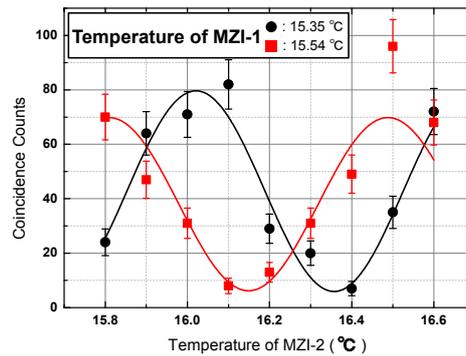

Fig.3. Two-photon interference fringes obtained after 300-km distribution over fiber. Black circles and red squares: experimental data when the temperature of MZI-1 was set at 15.35 and 15.54 °C, respectively. Statistical error bars are shown. The fitted curves were obtained without any statistical weighting of the experimental data.

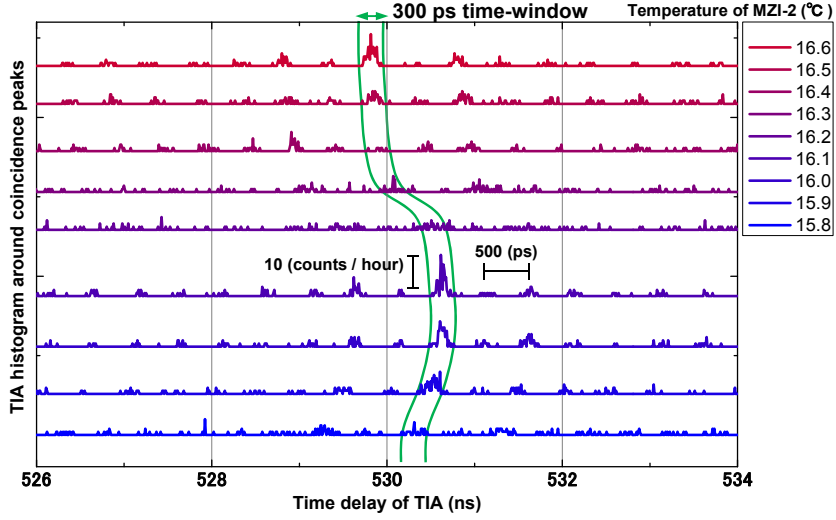

Fig.4. Fluctuations of temporal positions of coincidence peaks in TIA histograms. The temporal positions of the coincidence peaks were shifted during the measurement because of the fluctuations in fiber length caused by the unstable room temperature. The coincidence counts were collected within a 300 ps time window by tracing the coincidence peaks over a period of one hour.

Table 1. Coincidence counts for each phase setup

|  | $R(\theta_s, \theta_i)$ | $R(\theta_s, \theta_i+\pi)$ | $R(\theta_s+\pi, \theta_i)$ | $R(\theta_s+\pi, \theta_i+\pi)$ | $E(\theta_s, \theta_i)$ |
|---|---|---|---|---|---|
| $\theta_s = d_s, \theta_i = d_i$ | 158±14.5 | 44±9.8 | 27±8.9 | 163±14.7 | 0.64±0.074 |
| $\theta_s = d_s, \theta_i = d'_i$ | 164±14.7 | 46±9.9 | 36±9.4 | 158±14.5 | 0.59±0.071 |
| $\theta_s = d'_s, \theta_i = d_i$ | 140±13.9 | 44±9.8 | 44±9.8 | 169±14.9 | 0.56±0.070 |
| $\theta_s = d'_s, \theta_i = d'_i$ | 37±9.5 | 167±14.8 | 168±14.8 | 41±9.7 | −0.62±0.071 |

## 6. Discussion

Here we discuss possibility of realizing a longer distance entanglement distribution in the future. The distribution distance is limited in principle by the signal-to-noise ratio. As long as the dark count rate of the SSPD is sufficiently low, it is possible to attempt a longer distance entanglement distribution. Actually, the present dark count rate in our experimental setup of $3 \times 10^{-9}$ Hz within the 300 ps time window can satisfy the entanglement distribution requirement for a distance of 480 km. However, the measurement time is increased exponentially as the distribution distance increases. For example, the coincidence count rate should be reduced to 1 count per hour in the 400 km distribution by using our experimental setup. This means that the CHSH inequality measurement would take more than 66 days. An improvement in the single photon detector will help us to overcome this difficult situation. The detection efficiency of SSPDs has increased surprisingly quickly in the last several years [29]. We believe that the development of SSPDs with a near-unit detection efficiency while maintaining a small dark count rate and high timing resolution could be achieved in the near future. An SSPD with a quantum efficiency of over 90 %, for example, could reduce the time needed for a CHSH inequality measurement to 3 days, and the measurement time for each phase setting will be over 5 hours. On the other hand, long-term measurement requires a more stable experimental setup. In the present experiment, the fibers for both the signal and idler photons are placed in the same room (and are thus at the almost same temperature), and therefore most of the effects of fiber length fluctuations canceled each other out. Nevertheless, we observed fluctuations in the temporal positions of the coincidence peaks, which were approximately 250 ps/hour as shown in Fig. 4. With the 400 km distribution, the fluctuation of the temporal positions of the coincidence peaks can be larger than 500 ps, which is the time slot interval in our experiment. The situation becomes even worse in a field test: the fibers are implemented at different temperatures, and so the fluctuations in the temporal positions of the coincidence peaks become larger. Therefore, we need to implement a scheme to monitor the fiber length fluctuation with a short temporal interval. In short, the distribution distance is limited not only by the signal-to-noise ratio, but also by the measurement time and the stability of the experimental setup.

We can apply our entanglement distribution technology to entanglement-based QKD. A realistic approach for long distance QKD based on entanglement has been reported [30]. Here we estimated the performance of a QKD system based on the Bennett-Brassard-Mermine 1992 protocol [31] using our current setup based on a model described in [32]. The maximum distribution distance was calculated to be 520 km (260 km × 2) with a very low secure key rate of $1 \times 10^{-7}$ bit $s^{-1}$, which means that we need over 2700 hours to establish 1 key bit, and this is obviously useless in real communication. If we assume that we require a minimum secure key rate of 1 bit $s^{-1}$, the maximum distribution distance should be 220 km (110 km × 2). Thus, although QKD over 300 km in fiber is difficult to achieve with our present experimental setup, it will be possible with improved detectors and a more stable experimental setup for long measurement times.

## 7. Summary

We demonstrated the long distance distribution of time-bin entangled photon pairs over 300 km of optical fiber. For the low-noise generation of the entangled photon pairs, we employed the SPDC process in a PPLN waveguide with 2 GHz repetition pump pulses. The low dark count rate of the SSPD also helped to improve the signal-to-noise ratio of the system. We used a two-photon interference experiment after the long distance distribution to confirm that the correlation of the distributed photon pair was still preserved. We also confirmed the violation of Bell's inequality by measuring the value of $S = 2.41 \pm 0.14$ leading to the violation of 2.9 standard deviations with a mean photon number $\mu = 0.1$. This experimental result for entanglement distribution over 300 km of optical fiber illustrates the potential for fiber experiments related to long-distance quantum communication.

**Acknowledgement**

We thank Dr. M. Fujiwara of the National Institute of Information and Communications Technology (NICT) for lending us 200km-long DSFs.

References
1. N. Gisin and R. Thew, "Quantum communication," Nature Photonics **1**, 165 - 171 (2007).
2. S. Wang, W. Chen, J.F. Guo, Z. Q. Yin, H. W. Li, Z. Zhou, G. C. Guo, and Z. F. Han, "2 GHz clock quantum key distribution over 260 km of standard telecom fiber," Opt. Lett. **37**, 6, 1008 (2012).
3. A. Fedrizzi, R. Ursin, T. Herbst, M. Nespoli, R. Prevedel, T. Scheidl, F. Tiefenbacher, T. Jennewein, and A. Zeilinger, "High-fidelity transmission of entanglement over a high-loss free-space channel," Nature Phys. **5**, 389 (2009).
4. J. F. Dynes, H. Takesue, Z. L. Yuan, A. W. Sharpe, K. Harada, T. Honjo, H. Kamada, O. Tadanaga, Y. Nishida, M. Asobe, and A. J. Shields, "Efficient entanglement distribution over 200 kilometers," Opt. Express **17**, 11440 (2009).
5. J. Yin, J. G. Ren, H. Lu, Y. Cao, H. L. Yong, Y. P. Wu, C. Liu, S. K. Liao, F. Zhou, Y. Jiang, X. D. Cai, P. Xu, G. S. Pan, J. J. Jia, Y. M. Huang, H. Yin, J. Y. Wang, Y. A. Chen, C. Z. Peng, and J.W. Pan, "Quantum teleportation and entanglement distribution over 100-kilometre free-space channels," Nature **488**, 185 (2003).
6. X. S. Ma, T. Herbst, T. Scheidl, D. Wang, S. Kropatschek, W. Naylor, B. Wittmann, A. Mech, J. Kofler, E. Anisimova, V. Makarov, T. Jennewein, R. Ursin, and A. Zeilinger, "Quantum teleportation over 143 kilometres using active feed-forward," Nature **489**, 269 (2012).

7. M. Zukowski, A. Zeilinger, M. A. Horne, and A. Ekert, ""Event-Ready-Detectors" Bell Experiment via Entanglement Swapping," Phys. Rev. Lett. **71**, 4287 (1993).
8. H. J. Briegel, W. Dur, J. I. Cirac, and P. Zoller, "Quantum repeaters: the role of imperfect local operations in quantum communication," Phys. Rev. Lett. **81**, 5932-5935 (1998).
9. J.W. Pan, D. Bouwmeester, H. Weinfurter, and A. Zeilinger, "Experimental Entanglement Swapping: Entangling Photons That Never Interacted," Phys. Rev. Lett. **80**, 3891 (1998).
10. C. H. Bennett, G. Brassard, S. Popescu, B. Schumacher, J. A. Smolin, and W. K. Wooters, "Purification of Noisy Entanglement and Faithful Teleportation via Noisy Channels," Phys. Rev. Lett. **76**, 722 (1996).
11. W. Dür, H. J. Briegel, J. I. Cirac, and P. Zoller, "Quantum repeaters based on entanglement purification," Phys. Rev. A **59**, 169 (1999).
12. H. Takesue and K. Inoue, "Generation of polarization entangled photon pairs and violation of Bell's inequality using spontaneous four-wave mixing in a fiber loop," Phys. Rev. A **70**, 031802(R) (2004).
13. X. Li, P. L. Voss, J. Chen, J. E. Sharping, and P. Kumar, "Storage and long-distance distribution of telecommunications-band polarization entanglement generated in an optical fiber," Opt. Lett. **30**, 1201-1203 (2005).
14. I. Marcikic, H. de Riedmatten, W. Tittel, H. Zbinden, M. Legre, and N. Gisin, "Distribution of time-bin entangled qubits over 50 km of optical fiber," Phys. Rev. Lett. **93**, 180502 (2004).
15. H. Takesue, "Long-distance distribution of time-bin entanglement generated in a cooled fiber," Opt. Express **14**, 3453-3460 (2006).
16. H. Hubel, M. R. Vanner, T. Lederer, B. Blauensteiner, T. Lorunser, A. Poppe, and A. Zeilinger, "High-fidelity transmission of polarization encoded qubits from an entangled source over 100 km of fiber," Opt. Express **15**, 7853-7862 (2007).
17. T. Honjo, H. Takesue, H. Kamada, Y. Nishida, O. Tadanaga, M. Asobe, and K. Inoue, "Long-distance distribution of time-bin entangled photon pairs over 100 km using frequency up-conversion detectors," Opt. Express **15**, 13957-13964 (2007).
18. Q. Zhang, H. Takesue, S. W. Nam, C. Langrock, X. Xie, B. Baek, M. M. Fejer, and Y. Yamamoto, "Distribution of time-energy entanglement over 100 km fiber using superconducting single-photon detectors," Opt. Express **16**(8), 5776–5781 (2008).
19. C. Langrock, E. Diamanti, R. V. Roussev, Y. Yamamoto, M. M. Fejer, and H. Takesue, "Highly efficient single-photon detection at communication wavelengths by use of upconversion in reverse-proton-exchanged periodically poled $LiNbO_3$ waveguides," Opt. Lett. **30**, 1725-1727 (2005).
20. T. Honjo, H. Takesue, and K. Inoue, "Generation of energy-time entangled photon pairs in 1.5μm band with periodically poled lithium niobate waveguide," Opt. Express **15**(4), 1679–1683 (2007).
21. G. N. Gol'tsman, O. Okunev, G. Chulkova, A. Lipatov, A. Semenov, K. Smirnov, B. Voronov, A. Dzardanov, C. Williams, and R. Sobolewski, "Picosecond superconducting single-photon optical detector," Appl. Phys. Lett. **79**, 705–707 (2001).
22. J. F. Clauser, M. A. Horne, A. Shimony, and R. A. Holt, "Proposed Experiment to Test Local Hidden-Variable Theories," Phys. Rev. Lett. **23**, 880–884 (1969).
23. W. Tittel, J. Brendel, H. Zbinden, and N. Gisin, "Violation of Bell Inequalities by Photons More Than 10 km Apart," Phys. Rev. Lett. **81**, 3563–3566 (1998).
24. H. de Riedmatten, I. Marcikic, V. Scarani, W. Tittel, H. Zbinden, and N. Gisin, "Tailoring photonic entanglement in high-dimensional Hilbert spaces," Phys. Rev. A **69**, 050304 (2004).
25. J. D. Franson, "Bell inequality for position and time," Phys. Rev. Lett. **62**(19), 2205–2208 (1989).
26. S. Aerts, P. Kwiat, J. Å. Larsson, and M. Zukowski "Two-Photon Franson-Type Experiments and Local Realism," Phys. Rev. Lett. **83**, 2872–2875 (1999).
27. A. Aspect, J. Dalibard, and G. Roger, "Experimental Test of Bell's Inequalities Using Time- Varying Analyzers," Phys. Rev. Lett. **49**, 1804–1807 (1982).
28. P. G. Kwiat, K. Mattle, H. Weinfurter, and A. Zeilinger, "New High-Intensity Source of Polarization-Entangled Photon Pairs," Phys. Rev. Lett. **75**, 4337–4341 (1995).
29. F. Marsili, V. B. Verma, J. A. Stern, S. Harrington, A. E. Lita, T. Gerrits, I. Vayshenker, B. Baek, M. D. Shaw, R. P. Mirin, and S. W. Nam, "Detecting single infrared photons with 93% system efficiency," Nature Photonics **7**, 210–214 (2013).
30. T. Scheidl, R. Ursin, A. Fedrizzi, S. Ramelow, X. S. Ma, T. Herbst, R. Prevedel, L. Ratschbacher, J. Kofler, T. Jennewein, and A. Zeilinger, "Feasibility of 300km quantum key distribution with entangled states," New J. Phys. **11**, 085002 (2009).
31. C. H. Bennett, U. Brassard, and N. D. Mermin, "Quantum cryptography without Bell's theorem," Phys. Rev. Lett. **68**, 557-559 (1992).
32. B. Miquel and H. Takesue, "Observation of 1.5 μm band entanglement using single photon detectors based on sinusoidally gated InGaAs/InP avalanche photodiodes," New J. Phys. **11**, 045006 (2009).